\documentclass[prl,aps,twocolumn,showpacs,floatfix]{revtex4}
\usepackage{epsfig}
\usepackage{amssymb}
\usepackage{amsmath}
\usepackage{amsfonts}
\usepackage{epstopdf}
\usepackage{verbatim}
\usepackage[normalem]{ulem}
\usepackage{lipsum}

\usepackage{verbatim}

\usepackage{float}

\usepackage{graphicx}
\usepackage{color}
\usepackage{subcaption}
\begin{document}
\title {
Novel Energy Scale  in the Interacting 2D Electron System Evidenced  from
Transport and Thermodynamic
Measurements
}

\author{ L.\,A.~Morgun$^{1,2}$ \thanks {e-mail:
morgunl@gmail.com}, A.\,Yu.~Kuntsevich$^{1,2}$, and V.\,M.~Pudalov$^{1,2}$}

\address{$^1$ P.\,N.~Lebedev Physical Institute, 119991 Moscow, Russia\\
$^2$ Moscow Institute  of Physics and Technology, Moscow 141700, Russia}
\begin{abstract}

By analyzing the in-plane field magnetoconductivity, zero field transport, and thermodynamic spin magnetization in 2D correlated electron system in high mobility
Si-MOS samples, we have revealed a novel high energy scale $T^*$, beyond the Fermi energy.
In magnetoconductivity, we found a sharp onset of the novel regime $\delta \sigma(B,T) \propto (B/T)^2$ above a density dependent temperature $T_{\rm kink}(n)$,
the high-energy behavior that ``mimics'' the low-temperature diffusive interaction regime.  The zero field resistivity temperature dependence exhibits an inflection point $T_{\rm infl}$.
 In thermodynamic magnetization, the weak field spin susceptibility per electron, $\partial \chi /\partial n$ changes sign at $T_{dM/d n}$.
  All three notable  temperatures, $T_{\rm kink}$, $T_{\rm infl}$, and $T_{d M/ d n}$, behave critically $\propto (n-n_c)$,  are close to each other,  and are intrinsic to high mobility samples only; we therefore associate them with a novel energy scale $T^*$
caused by interactions in the 2DE system.

\end{abstract}
\maketitle

Two-dimensional  (2D) interacting low density carrier systems in the past two decades attracted considerable interest
\cite{kravreview_2001, pud_granada_2004}, demonstrating fascinating electron-electron interaction effects, such as metallic temperature dependence of resistivity
\cite{krav_1994, hanein_1998},  metal-insulator
transition (MIT) \cite{krav_1994, krav_1995}, strong positive magnetoresistance  (MR) in parallel field \cite{simonian_PRL_1997, pudalov-MR_JETPL_1997, shashkin-MR_PRL_2001, vitkalov-MR-insulator_PRB_2005}, strong renormalization of the effective mass and spin susceptibility \cite{pud_granada_2004, shashkin-m_PRB_2002}, etc.

Far away of the critical MIT density $n_c$, in the well ``metallic regime'', these effects are
explained within framework of the Fermi liquid theory - either in terms of interaction quantum corrections (IC) \cite{ZNA-R(T)_PRB_2001, ZNA_BPar_2001}, or
temperature dependent screening of the disorder  potential  \cite{gold-dolgopolov_PRB_1986, das-hwang,  das-hwang_PRB_2013, g-d-MR_JETPL_2000}.
Both theoretical approaches so far are used to treat the experimental  data on transport, and the former one -- also
  to determine the Fermi liquid coupling constants from fitting the transport and magnetotransport data to the IC theory.
In the vicinity of the critical region,  conduction is treated within renormalization group \cite{finkelstein_1984},
or Wigner-Mott approach  \cite{camjayi-dobro_natphys_2008, dobro-kotliar_WM_PRB_2012}. This regime is however  out of scope of the current paper.

On the other side,  numerous theories predict breakdown of the uniform paramagnetic 2D Fermi liquid state as interaction strength increases   \cite{
dharma_2003, narozhny_2000, khodel_2005, spivak, sushkov}.
However,  it remained almost unexplored how the potential instabilities may reveal themselves in thermodynamics and transport.

In this paper we report results  of the transport, magnetotransport and magnetization measurements with 2D correlated electron system, which reveal the existence of a novel characteristic energy scale $T^*$, that is smaller than the Fermi temperature $T_F$, but much bigger than $1/\tau$  (we set throughout the paper   $\hbar, k_B, \mu_B = 1$). Obviously, no such large energy scale may exist in the pure Fermi liquid. $T^*$ reveals itself (i) in the weak in-plane field magnetotransport,  (ii) in zero field transport, and (iii) in the spin magnetization per electron. In magnetoconductivity, we found a sharp onset of the novel regime $\delta \sigma(B,T) \propto (B/T)^2$ above a density dependent $T_{\rm kink}(n)$, the high-energy behavior that ``mimics'' the low-temperature diffusive interaction regime \cite{ZNA_BPar_2001}.  $T_{\rm kink}(n)$ correlates well with inflection point $T_{\rm infl}(n)$  in the zero field resistivity temperature dependence. Finally, the two remarkable temperatures
correlate  with the temperature $T_{dM/dn}$ for which the spin susceptibility per electron $\partial \chi /\partial n$ (and $\partial M/\partial n$) changes sign. All three notable  temperatures, $T_{\rm kink}$, $T_{\rm infl}$, and $T_{dM /dn}$,  behave critically $\propto (n-n_c)$, are intrinsic to high mobility samples only, and are close to each other; we therefore associate them with a novel energy scale $T^*$
caused by interactions in the 2DE system.

{\bf Experimental}.
The ac-measurements (5 to 17\,Hz) of resistivity were performed using the four-probe lock-in technique
 in magnetic fields up to $\pm 7\,T$. The  range of temperatures,   $0.4 - 20$\,K, was chosen so as to ensure the absence of the shunting conduction of bulk Si at
 the highest temperatures, and, on the low-temperature side,
  to exceed the valley splitting and intervalley scattering rate  \cite{valley_scattering}.  The studied high mobility (100)Si-MOS samples had $\approx 190$\,nm gate oxide
  thickness, and were
  lithographically defined as rectangular  Hall bars,  $0.8\times5$\,mm$^2$ \cite{samples}.
The magnetoconductivity measurements are performed similar to
Ref.~\cite{knyazev_JETPL_2006},
but in the much wider domain of densities and temperatures, from far above the MIT critical density and in the well-conducting regime $k_Fl \gg 1$ down to the
critical regime $k_Fl \sim 1$.

By rotating the sample with a step motor, we  aligned magnetic field in the 2D plane to within $1^\prime$ accuracy,  using the  weak localization
magnetoresistance as a sensor of the
perpendicular field component.
Carrier density $n$ was varied by the gate voltage $V_g$ in the range $(0.9-10)\times 10^{11}$\,cm$^{-2}$. The linear $n(V_g)$ dependence was determined from
quantum oscillation period measured in the perpendicular field orientation
during the same cooldown.

\begin{figure}[H]
\begin{center}
\includegraphics[width=240pt]{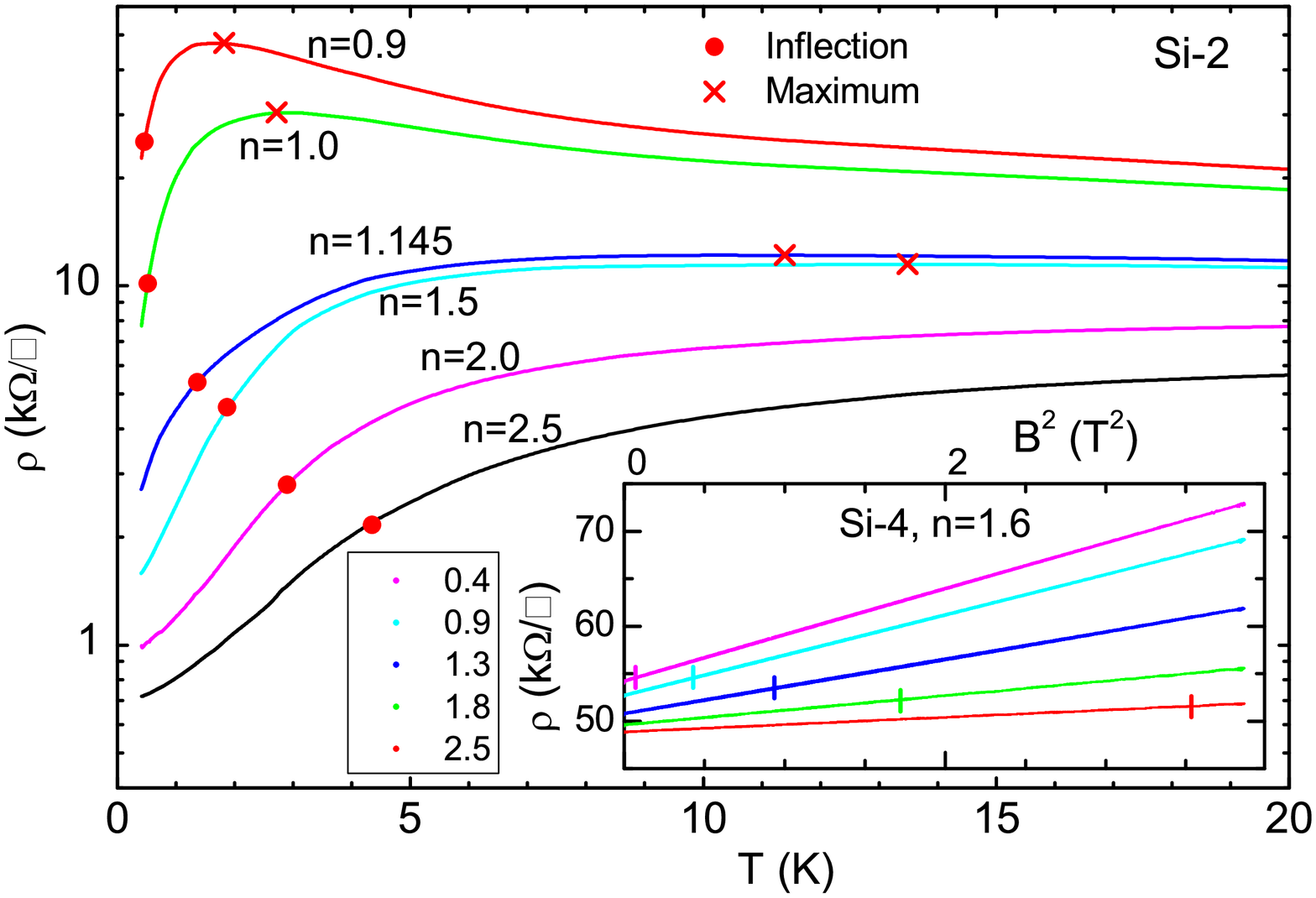}
\caption{(Color online) Temperature dependence of resistivity at zero field in the vicinity of $n_c$ ($\approx 0.85$) for sample
Si-2. The densities  are in $10^{11}$cm$^{-2}$.
Crosses mark the  $\rho(T)$ maxima, dots  -- the inflection points. The inset shows $\rho$ versus $B^2$ for sample Si-4 at a fixed
density and at five  temperatures:  0.4, 0.9, 1.3, 1.8, and 2.5\,K (from top to bottom). Vertical ticks mark $B=T$ field limit.}
\label{fig:rho(T)}
\end{center}
\end{figure}

%%%%%%%%%%%%%%%%%%%%%%%%%%%%%%%%%%%%%%%%%%%%%%%%%%%
{\bf In-plane field magnetoconductivity.}
The inset in Fig.~\ref{fig:rho(T)} shows that the
magnetoresistivity varies in weak fields
as $B^2$,   with a high accuracy.
From the $\rho(B)$ data
we determined the magnetoconductivity prefactor $a_\sigma = -\frac{1}{2} \partial^2\sigma/\partial B^2|_{B=0} \equiv (1/2\rho^2)\partial^2
\rho/\partial B^2 |_{B=0}$ which is analyzed below.

%%%%%%%%%%%%%%%%%%%%%%%%%%%%%%%%%%%%%%%%%%%%%%
%%%%%%%%%%%%%%%%%%%%%%%%%%%%%%%%%%%%%%%%%%%%

\begin{figure}[H]
\centering
\includegraphics[width= 230pt]{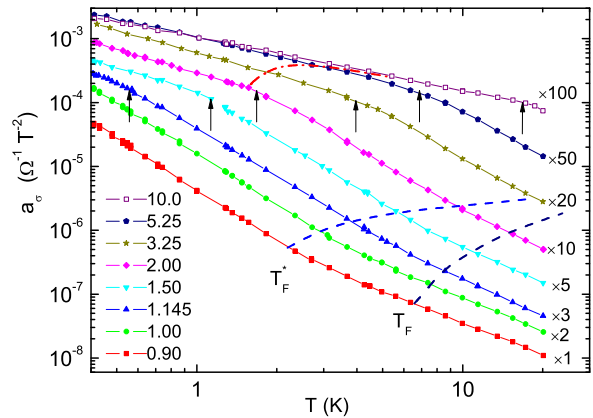}
\caption{(Color online) Temperature dependence of the magnetoconductivity prefactor $a_\sigma$  for sample Si-2, for eight electron densities  (increasing from
bottom to top),
in units of $10^{11}$cm$^{-2}$.  For clarity, the  curves are  magnified by the factors shown next to each curve.  Arrows mark  the kink positions, the dashed
curves show $T_F(n)$ and $T_F^*(n)$, dash-dotted curve shows $T_{\rm incoh}(n)$.
}
\label{fig:a_sigma}
\end{figure}

In the weak field limit, $B<T$, variations of the conductivity at a fixed temperature are low, $\leq 5\%$,  (see insert to Fig.~1). This smallness favors
comparison of the data with
theory of interaction corrections
which makes firm predictions
specifically for
magnetoconductivity (MC)
\cite{ZNA_BPar_2001}.
The magnetoconductivity  prefactor
$a_\sigma(T,n)$
is plotted in Fig.~\ref{fig:a_sigma} versus temperature.
In the wide density range, $(2 - 10)\times 10^{11}$cm$^{-2}$  in the well-conducting regime, the estimated diffusive/ballistic border \cite{ZNA-R(T)_PRB_2001},
$T_{\rm db} = (1 + F_0^\sigma)/2 \pi  \tau \approx 0.2$\,K, is below the accessible temperatures range of our measurements and we anticipate to observe the ballistic regime of interactions.

Surprisingly, at temperatures much higher than $T_{\rm db}$, the prefactor $a_\sigma(T)$ develops roughly $\propto T^{-2}$ (contrary to
the predicted ballistic-type dependence $a_\sigma \propto T^{-1}$) \cite{ZNA_BPar_2001}.
At somewhat lower temperatures (but still higher than $T_{\rm db}$),  the $a_\sigma(T)$ dependence
softens to $\propto T^{-1}$.
The crossover in
Fig.~\ref{fig:a_sigma} occurs rather sharply,  as a kink on the double-log scale.
The kink and the overall type of behavior is  observed in the wide range of densities and  for all
studied high mobility samples. Figure~3 shows that $T_{\rm kink}(n)$ develops critically versus electron density,
$\propto (n-n_c)$ where $n_c$ within experimental uncertainty coincides with MIT critical density.
The sharp crossover at high temperatures  to the novel regime of MC, which is in
contrast with the theory predictions, is one of the main results of our study. (In \cite{SOM} we compare the  magnetoresistivity and the magnetoresistivity prefactors).

%%%%%%%%%%%%%%%%%%%%%%%%%%%%%%%%%
%%%%%%%%%%%%%%%%%%%%%%%%%%%%%%%

{\bf Features in thermodynamics}.
In case the kink in magnetoconductivity indeed signals a novel energy scale, it must show up in temperature dependencies of other physical
quantities measured in the  high temperature range
 and in weak or zero magnetic fields.
Other available data which fits these requirements are as follows:
(i) spin magnetization per electron $\partial M/\partial n$, and
(ii) zero-field transport $\rho(T)$.

\begin{figure}[H]
\centering
\includegraphics[width=\linewidth]{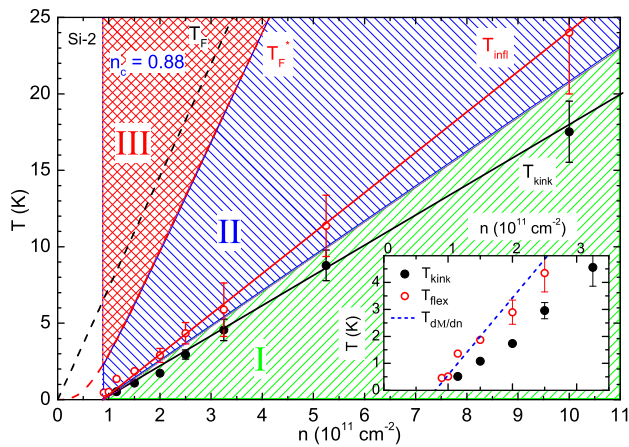}
\caption{ Empirical phase diagram of the 2DE system.
Dashed areas are: (I)
-- the ballistic interaction  regime, (II) --
novel MC regime. Hatched area (III) is the  non-degenerate regime, blank area -- localized phase. Full dots -- the kink  temperature $T_{\rm kink}$, open dots --
inflection point $T_{\rm infl}$.
Sample Si-2.  Dashed curves show the calculated  bare ($T_F$) and the renormalized ($T_F^*$) Fermi temperatures.
Insert blows up the low density region; dashed line is $T_{dM/dn}$ \cite{teneh_PRL_2012}.
}
\label{fig:phase diagram}
\end{figure}

The spin magnetization data \cite{teneh_PRL_2012} show a pronounced sign change of $\partial \chi/\partial n \equiv \partial^2 M/\partial B\partial n$ at a  density dependent temperature $T_{dM/dn}(n)$.
Physically,  the sign change means that for temperatures   lower than $T_{dM/dn}(n)$, the minority phase (large spin collective ``spin droplets'') melt as density increases. In other words, extra electrons
added to the system join the Fermi sea, improve screening and favor ``spin droplets'' melting.
For temperatures above $T_{dM/dn}(n)$, the number of ``spin droplets''  grows as density increases; here the extra electrons added to the 2D system prefer joining the ``spin droplets'' (see also \cite{SOM}).
 The  $T_{dM/dn}(n)$ dependence  copied from Figs.~1 and 2 of Ref.~\cite{teneh_PRL_2012} is depicted in the insert to  Fig.~\ref{fig:phase diagram}.
One can see that $T_{dM/dn}(n)$ behaves critically and vanishes to zero at $n_c$; remarkably,
within the measurements uncertainty, it is consistent with $T_{\rm kink}(n)$ deduced
from  magnetotransport.

{\bf Zero field transport}.
Figure \ref{fig:rho(T)} shows typical temperature dependencies of resistivity for high mobility Si-MOS samples. Each curve has two remarkable  points: the $\rho(T)$ maximum, $T_{\rm max}$, and
inflection,  $T_{\rm infl}$ \cite{knyazev_PRL_2008}.  Whereas $T_{\rm max}$ is an order of the renormalized Fermi energy, the  inflection point happens at much lower temperatures, in the degenerate regime.
Importantly, the inflection temperature
appears to be close to the kink temperature (see Figs.~\ref{fig:rho(T)}, \ref{fig:phase diagram}). Therefore, the proximity of the three notable temperatures which are inherent to high mobility samples solely,
$T_{\rm kink} \approx T_{\rm infl} \approx T_{dM/dn}$ strongly
suggests  the existence of a new energy scale $T^*$ in the correlated 2D system.
$T^*$ is much less than
the bare Fermi temperature $T_F$ \cite{ando_review}, and the renormalized  $T_F^* = T_F (m_b/m^*)$ \cite{pudalov_PRL_2002}; in contrast to $T_F$ (which is $\propto n$), $T^*(n)$ develops
as ($n-n_c$).
On the other hand, it  is much higher than   the ``incoherence'' temperature at which the phase coherence is lost (defined as $\tau_\varphi(T) =\tau$
 \cite{brunthaler_PRL_2001},
confirming that the kink, inflection and $\partial \chi/\partial n$ sign change are irrelevant to the single-particle coherent effects.

%%%%%%%%%%%%%%%%%%%%%%%%%%%%%%%%%%%%%%%%%%%%%%
%%%%%%%%%%%%%%%%%%%%%%%%%%%%%%%%%%%%%%%%%%%%

{\bf Phenomenological model for transport and magnetotransport}.
In the absence of an adequate microscopic theory, we attempt to elucidate the origin of the $T^*$ energy scale and of the novel magnetoconductance behavior.
We suggest below a phenomenological model that links ```high temperature" transport and magnetotransport behavior in a unified picture.

The features of our interest, $T_{\rm kink}$ and $T_{\rm infl}$
represent  ``high-energy'' physics.
Moreover, the  $\rho(T)$ (and $\sigma(T)$) variations of the experimental data
(Fig.~\ref{fig:rho(T)}) are so large, that the first order in $T$ corrections, of cause, cannot describe them.
Our analysis of other known theoretical models for homogeneous 2D Fermi liquid
\cite{pudalov_tbp} (see also \cite{SOM}) reveals that
neither of them describes  adequately the inflection on the $\rho(T)$ data and of cause does not include an associated energy scale. For this reason, we turn attention to
the two-phase state.

There is a large body of theoretical suggestions for spontaneous formation of the two phase state \cite{spivak,
shi-xie_2002, sushkov, narozhny_2000, dharma_2003} due to instabilities in the charge or spin channel, and many experimental indications
obtained with mesoscopic systems  or local probes \cite{ilani_Science_2001, ghosh_PRL_2004}. Finally, the spin magnetization measurements \cite{teneh_PRL_2012}
revealed the existence of the two-phase state of the macroscopic 2DE system, where the minority phase -- large spin droplets -- coexist with the majority Fermi liquid state.
Dealing with the two-phase state, the two channel scattering or additive resistivity approach seems quite adequate to the problem.

The typical functional form of the $\rho(T)$ (Fig.~1) also prompts dual channel scattering.
The simplest functional dependence that correctly describes  the inflection in $\rho(T)$ is provided by the phenomenological  form
 \cite{pudalov-SOI_JETPL_1997, pudalov_disorder_2001}:
\begin{equation}
\rho(T)=\rho_0 +\rho_1\exp\left(-\frac{\Delta(n)}{T}\right); \quad
 \Delta (n)= \alpha (n-n_c(B)),
\label{eq:rho-exp}
\end{equation}
where  $\rho_1(n,B)$ is  a slowly decaying function of $n$, and $\rho_0(n,T)$ includes Drude resistivity and quantum corrections, both from the single-particle
interference and interaction.
Although this model was suggested on a different ground,
it fits well resistivity data   in the vicinity of MIT for
various material systems (see \cite{SOM}). This simple additive   $\rho(T)$ form  satisfies general requirements for the transport behavior in the vicinity of a critical point \cite{amp_2001, knyazev_PRL_2008}, and also explains the apparent success of the earlier attempts of one-parameter scaling (namely of the $\rho(T)$ steep rise, mirror reflection, etc.)  \cite{krav_1994, krav_1995}.

Obviously, in this model $T_{\rm infl}= \Delta/2$. To take magnetic field into account, and following results of Refs.~\cite{pudalov_disorder_2001, vitkalov-form}
we include to $(\Delta/T)$ all the lowest order in $B/T$ (and even-in-$B$)  terms, as follows:
\begin{equation}
\Delta(T,B,n)/T =\Delta_0(n)/T - \beta(n) B^2/T  - \xi(n) B^2/T^2,
\label{eq:nc(B)}
\end{equation}
with $\Delta_0=\alpha[n-n_c(0)]$.

 Eqs.~(\ref{eq:rho-exp}) and (\ref{eq:nc(B)})
 link the magnetoconductance  with the zero-field $\rho(T)$
 temperature  dependence. With these, the $\rho(T,B)$ dependence is as follows:
\begin{eqnarray}
\rho(B,T) &=& \left[\sigma_D -
\delta\sigma \cdot \exp \left(- T/T_B \right)\right]^{-1} \nonumber \\
&+ &\rho_1 \exp \left( - \alpha \frac{n-n_c(0)}{T} - \beta \frac{B^2}{T} - \xi \frac{B^2}{T^2}\right).
\label{eq:r(B&n)}
\end{eqnarray}

The term in the square brackets includes the Drude conductivity and interaction quantum corrections \cite{ZNA-R(T)_PRB_2001,
ZNA_BPar_2001}. The latter, $\delta\sigma(T)= \gamma (B^2/T) + \eta T$,  was calculated using experimentally determined $F_0^\sigma (n)$ values
\cite{pudalov_PRL_2002, klimov_PRB_2008}, and $\sigma_D$ found from a standard procedure
\cite{pudalov_R(T)_PRL_2003}. In
order to cut-off the corrections above a certain border temperature \cite{cut-off}  and, thus, to disentangle the exponential- and linear-in-$T$  contributions,
the calculated interaction correction are cut-off with an exponential crossover function above $T_{\rm B} $ which for simplicity we set equal to $\Delta(n)/2$.

\begin{figure}[H]
 \centering
\includegraphics[width=\linewidth]{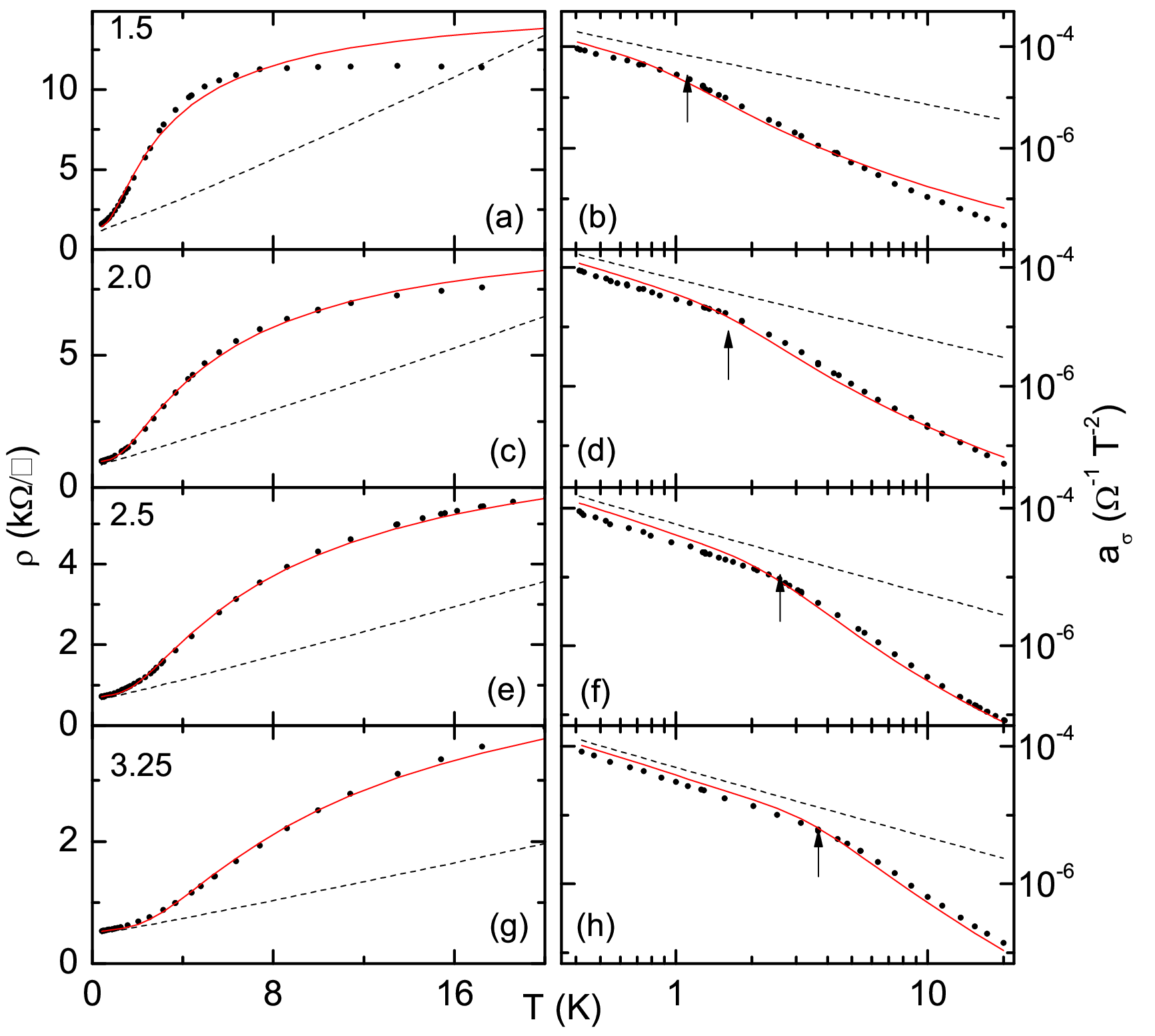}
\caption{Fitting $\rho(T,B=0)$ dependencies (left panels) and  $a_\sigma(T)$ (right panels) with the same set of the fitting parameters. Sample
Si-2; carrier densities (from
top to bottom) are $n=1.5; 2.0; 2.5,$ and $3.25 \times 10^{11}$\,cm$^{-2}$. Fitting parameters are presented in the Table. }
\label{fig:dualFit}
\end{figure}

From Eq.~(\ref{eq:r(B&n)}),
 the prefactor  $a_\sigma = - (1/2)\partial^2\sigma/\partial B^2 $ is calculated straightforward and
in Fig.~\ref{fig:dualFit}
is compared with experimental data.
In the $\rho(T)$ fitting [Figs.~4\,(a,c,e,g)],  basically, there is only one adjustable parameter,  $\rho_1(n)$,  for each density. Indeed,
$n_c(0)$ is determined from the conventional scaling analysis at $B=0$ \cite{knyazev_PRL_2008}, and the slope, $\alpha = 2 \partial T_{\rm infl}(n)/\partial n$
may be determined  from Fig.~\ref{fig:phase diagram}. However, in order to test the assumed linear $\Delta(n)$ relationship, Eq.~(1), we treated  $\alpha(n)$ as an adjustable
parameter. On the next step, in the $a_\sigma(T)$ fitting [Figs.~4\,(b,d,f,h)] we fixed the parameters determined from the $\rho(T)$ fit, and
varied  $\beta(n)$ and $\xi(n)$.

One can see that both $\rho(T)$ and $a_\sigma(T)$ are well fitted; the model captures correctly the major data features, the steep $\rho(T)$  rise (including the
inflection), and the $a_\sigma(T)$ kink. Within this  model, the kink signifies a transition from the low-temperature magnetoconductance regime of ballistic
interaction (where the
exponential term may be neglected and where the quantum corrections dominate) to the high temperature regime governed by the steep exponential $\rho(T)$ rise. The
parameters of the fit (Figs.~\ref{fig:dualFit}) are summarized in the Table of the Supplementary materials. The factor $\beta$ is an order of magnitude smaller than $\xi$, therefore, the corresponding
term in Eq.~(3) becomes important only at high temperatures. The slope, $\alpha$, is almost constant, confirming our assumption  [(Eq.~(1)].

{\bf Impact on the MC interpretation}.
For high densities $n\gg
n_c$,
the temperature  range above $T_{\rm kink}$
is unambiguously beyond the diffusive regime of interactions and, hence,  the  $B^2/T^2$ dependence is
the novel high-temperature MC regime of the {\em non-diffusive type}.
Below $T_{\rm kink}$ the temperature is still higher than $T_{\rm db}$ and  the regime $a_\sigma \propto T^{-1}$ (see Fig.~\ref{fig:a_sigma})
therefore is reminiscent of the  standard ballistic interaction regime \cite{ZNA_BPar_2001}. This conclusion is confirmed by Figs.~4b,d,f,h
where the standard interaction corrections (incorporated in Eq.~3) with experimentally determined interaction parameters provide quite a sucsessful fit below $T_{\rm kink}$.

The two-stage
``high-temperature'' dependence of the magnetoconductivity prefactor is universal and persists from well conducting regime far above the critical density,  down to the
 very critical density.
As a result, in the vicinity of MIT, the ``high-temperature'' MC regime $\delta\sigma \propto -(B/T)^2$,
mimics the  behavior anticipated for the {\em diffusive regime} of quantum interaction
\cite{ZNA_BPar_2001, CdCL_PRB_1998}.
This fact  therefore casts serious doubt on the RG  treatment of the experimental $\rho(T,B_\parallel)$ data in the vicinity of MIT, and particularly, on the
phase diagram  of 2D interacting and disordered system deduced from fitting the experimental data with the RG-theory \cite{punnoose_PRB_2010, anissimova_nphys_2007}.

{\bf Conclusion}.
To conclude, we have found a novel high energy scale $T^*$ to exist in the correlated 2D electron system, beyond the Fermi energy.
It reveals itself in transport, in-plane field magnetotransport and thermodynamics.
All three
notable temperatures behave critically, $\propto (n-n_c)$, and are rather close to each other.
For temperatures above the density dependent $T^*$, the in-plane field MC  crosses over from the conventional ballistic-type $-(B^2/T)$
 dependence to the novel $-(B^2/T^2)$ dependence.
We suggested phenomenological description of the transport and magnetotransport data, based on the two phase state (two resistivity channels).

Our  present studies are performed with high mobility Si-MOS samples which show well pronounced MIT.  Therefore, the results obtained send a warning to
interpretation of the in-plane field MC in the vicinity of MIT, particularly within the framework of the RG theory~\cite{pf_science_2005}.  Secondly, our results explain why the Fermi-liquid  parameters
extracted from fitting the measured  MC scatter
significantly in various experiments: indeed, by  fitting the data in the nominally ballistic regime,  one would observe $a_\sigma$ (and deduce $F_0^a$ values) strongly dependent on
the particular temperature range, above or below the kink.

Clearly, there is a need in a microscopic theory that must explain on the same footing all three critical behaviors:
in the zero field resistivity,
in magnetoconductivity,
and in spin susceptibility per electron.
A possible origin of the $T^*$ scale may be such a  structure of collective energy levels for individual droplets of the minority phase,
which in analogy with quantum dots may cause features simultaneously in  thermodynamics and in transport of itinerant electrons.

\section{Acknowledgements}
We thank I.\,S.~Burmistrov, I.~Gornyi, and A.\,M.~Finkel'stein for discussions. The magnetotransport  measurements were supported by RFBR (12-02-00579),
transport and thermodynamic measurements - by Russian
Science Foundation (14-12-00879). The work was done using research equipment of the Shared Facilities Center at LPI.


\begin{thebibliography}{99}
\bibitem{samples}  Measurements were performed with three  high mobility samples, Si-2, Si63, and Si-4  ($\mu^{\rm max} = 3$, 2.5, and 0.95\,m$^2$/Vs)
which showed similar behavior in transport and magnetotransport (with slightly different $n_c$ values).
In contrast, for two low mobility samples Si-40 and  Si-46 ($\mu^{\rm max} \approx 0.2$ and 0.1m$^2$/Vs)
the magnetoconductivity  behaved in accord with the IC theory.
\end{thebibliography}
\end{document}

\title{SUPPLEMENTARY ONLINE MATERIALS to: Novel Energy Scale  in the Interacting 2D Electron System Evidenced  from Transport and Thermodynamic Measurements}

\author{
L.\,A.~Morgun$^{1,2}$,
A.~Yu.~Kuntsevich$^{1,2}$,
V.~M.~Pudalov$^{1,2}$
}
\affiliation{$^1$ Lebedev Physical Institute of the RAS, 119991 Moscow, Russia\\
$^2$ Moscow Institute of Physics and Technology, 141700 Moscow, Russia
}

\pacs{71.30.+h, 73.40.Qv, 71.27.+a}
\maketitle

\subsection{Comments on
calculation of $a_\sigma$ from the measured $\rho(B)$ dependence}

 In weak in-plane fields:
\begin{align}
\label{sigma}
\sigma &= \sigma_0 - a_\sigma  B^2 + o\left(B^2\right) \\
\label{rho}
\rho &= \rho_0 + a_\rho B^2 + o(B^2),
\end{align}
 where  $B$ is considered to be small as compared with either $T$, or $(T^2\tau)$  ($\hbar$, $k_B$ and $\mu_B$ are set to unity throughout the paper, for shortness), and
by definition
\begin{eqnarray}
a_\sigma &\equiv & \left. -\frac{1}{2} \partial^2\sigma/\partial B^2 \right|_{B=0} \nonumber \\
a_\rho &\equiv & \left. \frac{1}{2} \partial^2\rho/\partial B^2 \right|_{B=0} \nonumber,
\end{eqnarray}

Consider the relation between $a_\sigma$ and the experimentally measured $\rho(B)$.
In purely parallel magnetic field  $\sigma = 1/\rho$.

Taking the second derivative  from both sides of equation (\ref{sigma}) and recalling that
$\left. \left(\partial \rho/\partial B\right) \right|_{B=0} = 0$, we obtain:
\begin{equation}
a_\sigma = \left[\frac{1}{2 \rho^2} \frac{\partial^2 \rho}{\partial B^2} - \frac{1}{\rho^3} \left(\frac{\partial \rho}{\partial B} \right)^2\right]_{B=0} = \frac{1}{2 \rho^2} \frac{\partial^2 \rho}{\partial B^2}
\end{equation}
The latter proves  Eq.~(1) used in our paper to calculate $a_\sigma$ from the measured $\rho(B)$ dependence.

In the experimental data,  purely parabolic $\rho(B)\propto B^2$ dependence was found to extend with high accuracy far above the range of low fields ($B<T$) which we used  for calculating  $a_\sigma$. For example, the
higher order-in-$(B/T)$ terms in Eqs.~(\ref{sigma}) were less than  0.1\%  (relative to the $B^2$-term) even at $B/T = 6.5$, and have been neglected therefore for $B\ll T$.

\subsection{Magnetoconductivity and  magnetoresistivity}
Variations of the conductivity with weak in-plane field at a fixed temperature are low, $\leq 5\%$, in the selected range of fields $B<T$ (see insert to Fig.~1 of the main text). This smallness enables to apply the theory of interaction corrections (IC) which makes firm predictions for the  functional temperature dependence of the weak field magnetoconductivity (MC) and suggests a clear physical picture behind it \cite{ZNA_BPar_2001}.
Therefore, in the main text of our paper
we analyze namely the MC prefactor, $a_\sigma =-(1/2) \partial^2\sigma/\partial B^2|_{B=0}$ versus $T$. In Fig.~1a here and in the main text, one can see that $a_\sigma$ develops $\propto T^{-1}$ in the low-temperature part of the ballistic regime ($(1+F_0^a)/2\pi\tau \approx 0.2$\,K $<T$), in a qualitative agreement with the IC predictions. However, for higher temperatures, still in the ballistic and, presumably, degenerate regime $T<T_F$, there is a sharp kink   and the onset of the novel unforseen dependence, $a_\sigma(T) \propto T^{-2}$.

This effect is observed for several high and moderate mobility Si-MOS samples (see Ref. [26] in the main text), and is missing in more disordered (a factor of 10 lower mobility) samples, where $a_\sigma(T)$ dependence is qualitatively consistent with theory of interaction corrections. Importantly, by contrast to the Fermi energy which in 2D  is  proportional to the carrier density, the novel energy  scale develops critically, $\propto (n-n_c)$, with slightly sample mobility dependent $n_c$ values. It vanishes to zero as density approaches the critical value (see Fig.~3 of the main text); these two facts confirm the relevance of the electron-electron interaction effects.
Another indication of the crucial importance of the electron-electron interactions is the fact that the kink in $a_\sigma(T)$ at $T_{\rm kink}$ and the novel regime of MC at $T>T_{\rm kink}$ are intrinsic only to high mobility samples, where the strongly correlated regime is accessed  upon lowering density. For samples Si-40 and Si-46 with a factor of 10 -- 30 lower mobilities, in the same range of temperatures, the magnetoconductance  develops in accord with  IC theory predictions for the diffusive regime, with no kink.

The magnetoconductivity in the in-plane field is inequivalent to magnetoresistivity (MR), because variations of the conductivity with temperature at zero field are large (a factor of 4 -- 7) in high mobility Si-MOSFETs. As a result,  the $a_\rho (T)=(1/2)\partial^2\rho/\partial T^2|_{B=0}$ temperature dependence is different from that for $a_\sigma(T)$ (see Fig.~1b): it is less transparent, being affected by both, the onset of the novel regime in MC and by the strong $\rho(T)$ (and $\sigma(T)$) dependencies.
For higher densities, where the $\rho(T)$ variations are relatively weak (see the lowest 4 curves in Fig.~1b),  $a_\rho(T)$ exhibits a maximum that coincides with the kink in $a_\sigma(T)$ for densities $n=10, 5.25, 3.25, 2.5\times 10^{11}$\,cm$^{-2}$. For lower densities the maxima in $a_\rho(T)$  get smeared which hampers their quantifying.
The simplicity of the $a_\sigma(T)$ dependence (in comparison with $a_\rho (T)$)
clearly points at the  primary role of the
magnetoconductivity rather than magnetoresistivity for the studied system.

The kink temperature  $T_{\rm kink}$ in $a_\sigma$ lies far away from the bare and renormalized Fermi energy  and from $(1+F_0^a)/2\pi\tau \approx 0.2$\,K value \cite{ZNA-R(T)_PRB_2001} which are the only known energy scales in the Fermi liquid. We interpret $T_{\rm kink}$  as a manifestation of an additional energy scale, beside the Fermi energy. Obviously, no such energy
scale may exist  in the pure 2D Fermi liquids, and vice versa, its existence  indicates a non-Fermi liquid state.

 In Fig.~1a, one can also see that the magnetoconductivity prefactor exhibits another twist upward, clearly noticeable for the four lowest curves (lowest densities). However, this feature occurs at approximately renormalized $T_F$
and is likely to signify  a transition to non-degenerate regime, which is beyond the scope of our paper.

%%%%%%%%%%%%%%%%%%%%%%%%%%%%%%%%%%%%%%%%%%%%%%
\begin{figure}[h]
\begin{center}
\includegraphics[width=240pt]{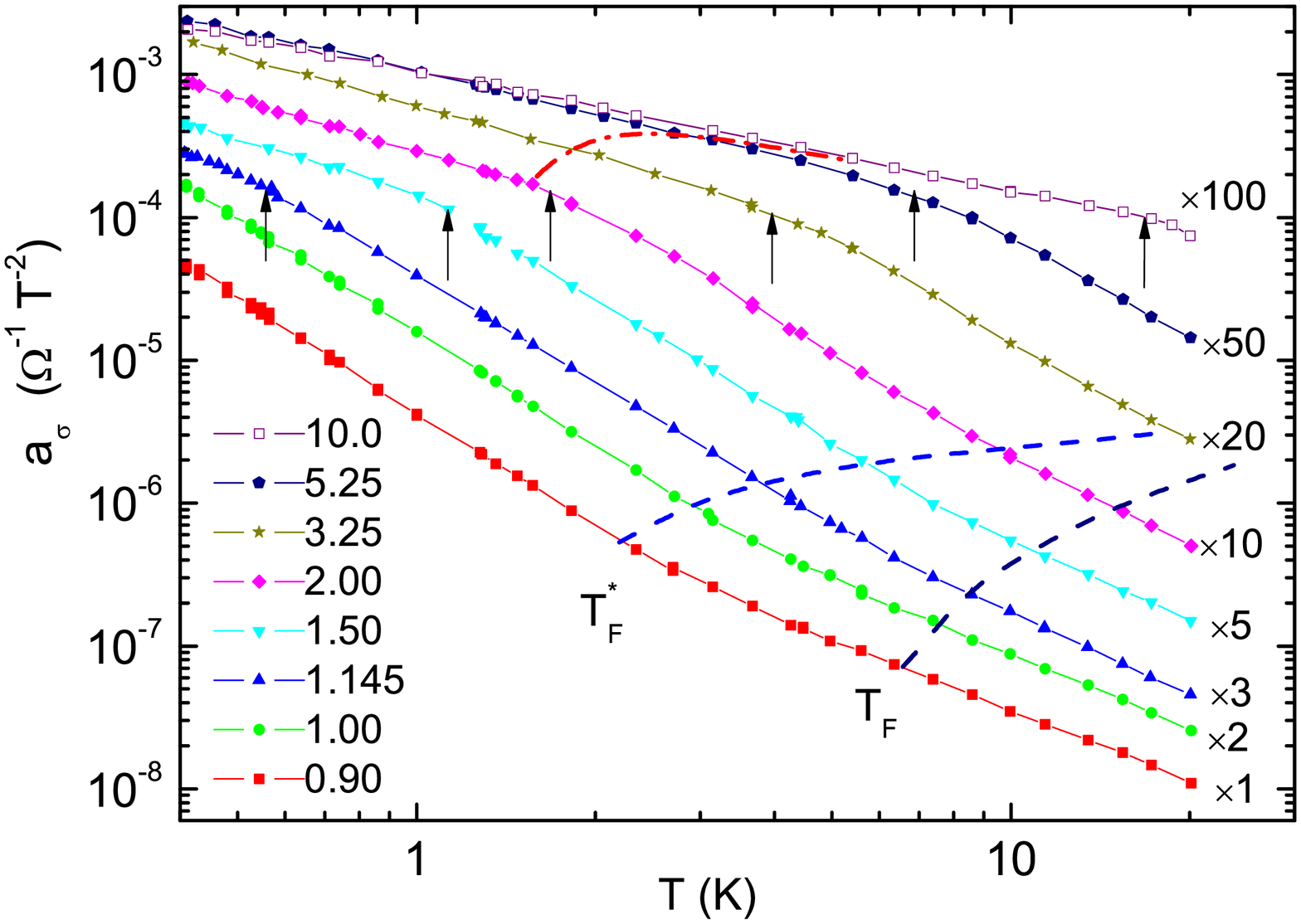}
\includegraphics[width=240pt]{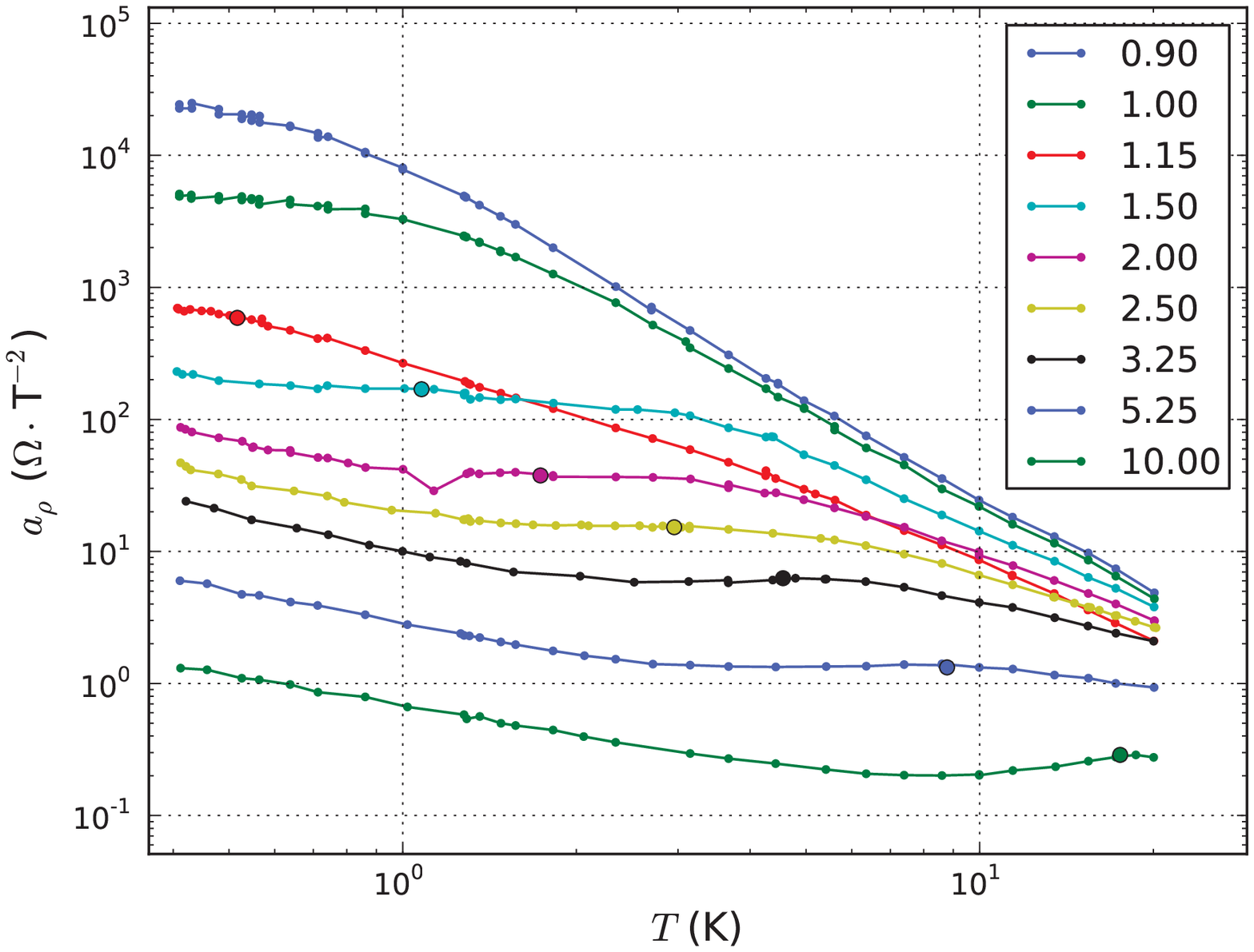}
\caption{(Color online) Temperature dependencies  of (a) $a_\sigma(T)$ and (b) $a_\rho$. Arrows (left panel) and dots (right panel) mark $T_{\rm kink}$ - kink temperature in $a_\sigma$. }
\label{fig:arho(T)}
\end{center}
\end{figure}
%%%%%%%%%%%%%%%%%%%%%%%%%%%%%%%%%%%%%%%%%%%%

\subsection{Other available data:
spin magnetization}

%%%%%%%%%%%%%%%%%%%%%%%%%%%%%%%%%%%%%%%%%%%%%%
\begin{figure}[h]
\begin{center}
\includegraphics[width=250pt]{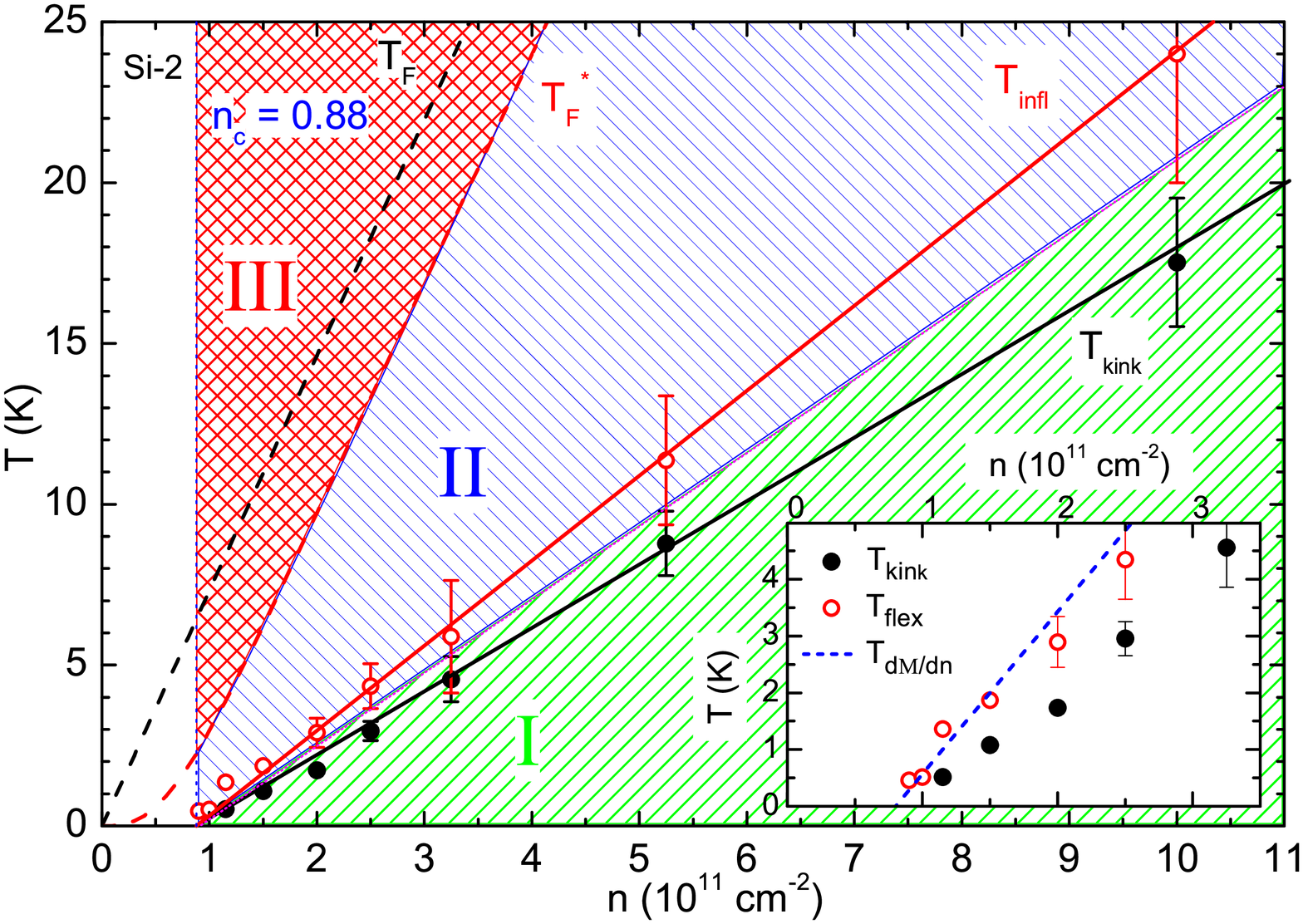}
\caption{Empirical phase diagram of the 2DE system.
Dashed areas are: (I) --  ballistic interaction  regime, (II) -- novel MC regime.
Hatched area (III) is the  non-degenerate regime, blank area -- localized phase. Full dots -- the kink  temperature $T_{\rm kink}$, open dots --
inflection point $T_{\rm infl}$. %on the $\rho(T)$  curves.
Sample Si-2.  Dashed curves show the calculated  bare ($T_F$) and the renormalized ($T_F^*$) Fermi temperatures.
Insert blows up the low density region; dashed line is $T_{dM/dn}$ -- sign change of $dM/dn$ \cite{teneh_PRL_2012}.
}
\label{fig:phase digram}
\end{center}
\end{figure}
%%%%%%%%%%%%%%%%%%%%%%%%%%%%%%%%%%%%%%%%%%%%

In order to test whether the kink temperature  in magnetoconductivity has a more general  significance,
we inspected temperature dependencies of other physical quantities  in the  high temperature range
 and in weak or zero magnetic fields.
Other available data which fit these requirements are as follows:\\
(i) spin magnetization per electron $\partial M/\partial n$ \cite{teneh_PRL_2012}, and\\
 (ii) zero-field transport $\rho(T)$.

 The spin magnetization-per-electron $\partial M/\partial n$  data \cite{teneh_PRL_2012}, in general,
 are interpreted as a clear evidence for the formation of a two-phase state, in which the Fermi liquid phase  coexists with large spin collective ``spin droplets'' (the latter being presumably collective localized states).
 These $\partial M/\partial n$ data, in particular,  show a pronounced sign change of $\partial \chi/\partial n \equiv \partial^2 M/\partial B\partial n$ at a  density dependent temperature $T_{dM/dn}(n)$.
Physically,  the positive sign
of the spin susceptibility per electron, $\partial \chi/\partial n$, means   that for temperatures  higher than $T_{dM/dn}(n)$,  the number of ``spin droplets''  grows as density increases.
 Here the extra electrons added to the 2D system prefer joining the ``spin droplets'', increase its magnetization $M$ and cause positive sign of $\partial \chi/\partial n$.

For temperatures lower than  $T_{dM/dn}(n)$, the number of ``spin droplets'' (minority phase)  decreases as density increases, i.e.
the  ``spin droplets'' melt as density increases. In other words, extra electrons
added to the system join the Fermi sea, improve screening and favor spin droplets ``melting'' that
results in the  negative sign of $\partial \chi/\partial n$.
 The  $T_{dM/dn}(n)$ dependence  copied from Figs.~1 and 2 of Ref.~\cite{teneh_PRL_2012} is depicted in the insert to  Fig.~\ref{fig:phase digram}.
One can see that $T_{dM/dn}(n)$ also behaves critically and vanishes to zero at $n_c$; remarkably,
within the measurements uncertainty, it is  consistent with $T_{\rm kink}(n)$ deduced
from  magnetotransport.

\subsection{Other available data: resistivity and conductivity in zero field}
Searching for  manifestation of the novel energy scale in zero-field transport
we analyze the  $\rho(T)$ and $\sigma(T)$ dependencies at zero field. The variations of these quantities in the relevant temperature range are  large (up to a factor of  7), making the IC theory inapplicable in this ``high temperature'' regime.
One can see from  Fig.~3a below that the $\rho(T)$ temperature dependence is  monotonic up to the limits of degeneracy, $T=T_F$, and follows one and the same additive resistivity functional form (Eq.~(1) of the main text) over a wide density range:
\begin{eqnarray}
\rho(T)&=&\rho_0 +\rho_1\exp(-\Delta(n)/T),\nonumber \\
 \Delta (n)&=& \alpha (n-n_c(B)),
\label{eq:rho-exp}
\end{eqnarray}
where  $\rho_1(n,B)$ is  a slowly decaying function of $n$, and $\rho_0(n,T)$ includes Drude resistivity and quantum corrections, both from the single-particle interference and interaction.

By inspecting available in literature data by several research groups  for
$\rho(T)$ in the low density correlated regime, one finds  that  the above empirical  additive resistivity form   fits well the $\rho(T)$ dependence for a number of  material systems \cite{pudalov_JETPL_1997a, hanein_1998, papadakis_1998, gao_2002, brunthaler_SOI_2010}. The additive resistivity form (cf. Matthiessen's rule)  naturally corresponds to the two-phase state of the low density 2D electronic system. The two-phase state is experimentally  revealed in macroscopic magnetization measurements \cite{teneh_PRL_2012}, local probe experiments \cite{ilani_Science_2001}, and  also widely discussed in theory \cite{spivak, sushkov, dharma_2003, shi-xie_PRL_2002, narozhny_2000, khodel_2005}.
This empirical  additive $\rho(T)$ form  satisfies general requirements for the transport behavior in the vicinity of a critical point \cite{amp_2001, knyazev_PRL_2008}, and explains the apparent success of the earlier attempts of one-parameter scaling (namely of the $\rho(T)$ steep rise and the mirror reflection symmetry between $\rho(T)$ and $\sigma(T)$ on the metallic and insulating side of the MIT)  \cite{krav_1994, krav_1995}.

In a close vicinity of the critical density ($n_c=0.85\times 10^{11}$cm$^{-2}$ for  sample Si-2), $\rho(T)$ becomes non-monotonic, with a slow decay after passing through a maximum. The maximum and the subsequent decay are explained as transition to the non-degenerate regime and subsequent increasing the scattering rate due to increase in the available phase space for the scattered electrons \cite{dassarma_2000}. The non-degenerate regime is however beyond the scope of out study.

%%%%%%%%%%%%%%%%%%%%%%%%%%%%%%%%%%%%%%%%%%%%%%
\begin{figure}[h]
\begin{center}
\includegraphics[width=240pt, height=190pt]{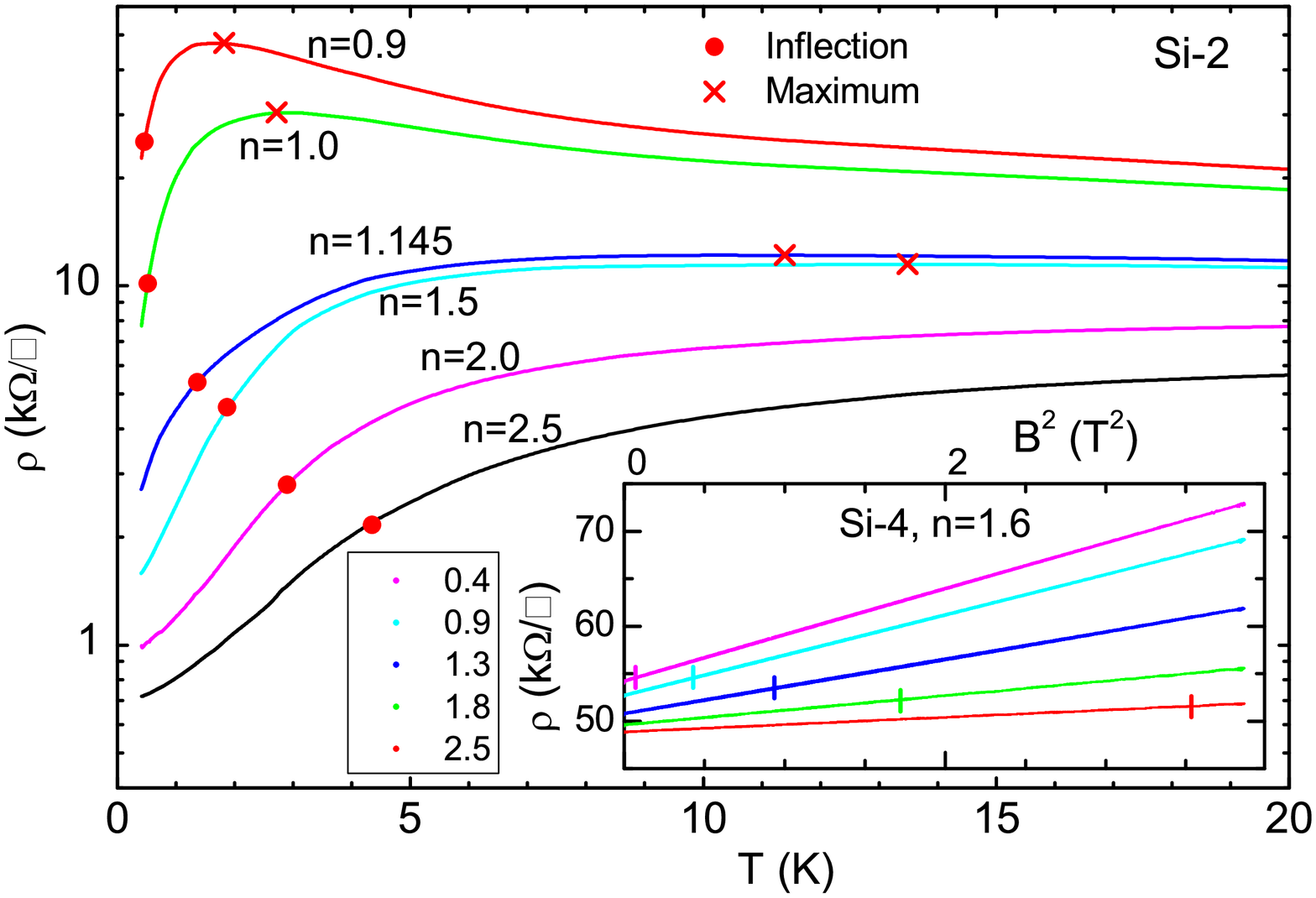}
\includegraphics[width=240pt, ]{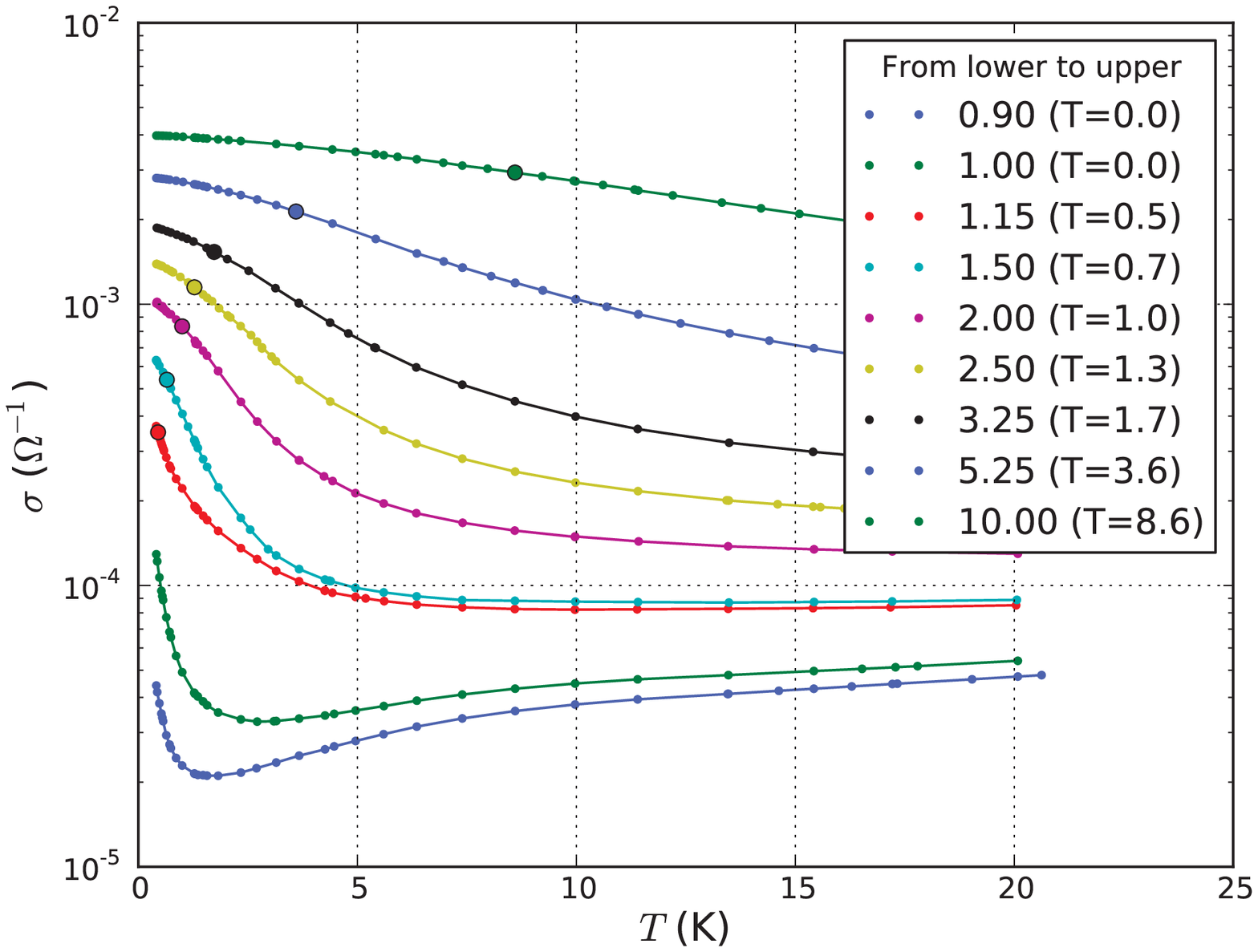}
\caption{(Color online)  Zero field temperature dependencies of (a) $\rho(T)$, and (b) $\sigma(T)$. Sample Si-2. }
\label{fig:sigma(T)}
\end{center}
\end{figure}
%%%%%%%%%%%%%%%%%%%%%%%%%%%%%%%%%%%%%%%%%%%%

Following the ancient maxim ``The simple is the seal of the true'',  we conclude on the primary role of the {\it scattering rate} (i.e. two channel scattering) rather than scattering time
(i.e., two conductivity channels) in this ``high-temperature''  transport in the two-phase metallic-like state at zero field.

In accord with above arguments, we focus further on the $\rho(T)$ dependence (shown in Fig.~1 of the main text and also in Fig.~3 here). On the generic $\rho(T)$ experimental curve there are two remarkable points: the maxima and inflection points,
both behave critically and vanish to zero at $n=n_c$ \cite{knyazev_PRL_2008}. The  temperature of the maximum is much higher than $T_{\rm kink}$ and is close to the nondegeneracy, whereas the inflection point $T_{\rm infl} \ll T_F$ appears to be  very close to $T_{\rm kink}$.
The inflection   and large $\rho(T)$ variation is known to be inherent to high mobility samples \cite{maunterdorf, varenna, pudalov_JETPL_98}; as mobility decreases (and, hence, critical density increases, and interaction  weakens) the magnitude of the $\rho(T)$ variation diminishes.
We again conclude that this feature (inflection) is  related with electron-electron interactions.

To summarize, all three notable temperatures, $T_{\rm kink}(n)$,  $T_{\rm infl}(n)$, and $T_{dM/dn}$ (i) are close to each other, (ii) behave critically, and (iii) are intrinsic to clean samples/strong interactions regime solely.
These arguments support our suggestion that they indicate a novel ``high energy'' scale $T^*(n)$  in the two-phase strongly correlated state.

\subsection{Comparison of the zero field transport with theory}

The interaction corrections theory, being the low-energy  theory,  does not include an additional energy scale $T^*$.
On the other hand, within the framework of the temperature dependent screening model,
the large $\rho(T)$ raise,  its further saturation, and  decay (for low densities) are explained  by smearing of the Fermi distribution function in the scattering time averaged in the energy space, that produces negative $d\rho/dT$, i.e. by a transition from the degenerate to nondegenerate regime above $T_F$ \cite{dassarma_2000}.
Within the screening theory  the low temperature analytical expansion of  resistivity is given by \cite{gold-dolgopolov_PRB_1986, das-hwang_PRB_2003, das-hwang_metallicity_PRB_2004}:
 \begin{equation}
 \frac{\delta\rho}{\rho} = C_1\left(\frac{T}{T_F}\right) + C_{3/2} \left(\frac{T}{T_F}\right)^{3/2} -B_2\left(\frac{T}{T_F} \right)^2,
 \label{screening R(T) expansion}
 \end{equation}
 where all prefactors are positive,
 $$
  C_1=\frac{2}{1+\frac{1}{q_0f}}, \quad C_{3/2}=\frac{2.646}{(1+\frac{1}{q_0f})^2}, \quad  B_2=\frac{\pi^2p(p+1)}{6},
  $$
$q_0=q_{TF}/2k_F\propto n^{-1/2}$, and $f=f(2k_F)$ is the
quasi-2D subband form-factor at the wave vector
$2k_F$ ($f=1$) in with the strictly 2D limit.

 The inflection in this analytical expansion arises due to the  third term.  In the $T/T_F\ll 1$  limit, the inflection point   $T^*/T_F \sim (1-0.12\sqrt{n/10^{11}})$ \cite{das-hwang_PRB_2003, das-hwang_metallicity_PRB_2004}, however {\it decreases}, rather than grows with density. From more exact numerical calculations, in Fig.~3b of Ref.~\cite{das-hwang_PRB_2013}, the inflection point also {\it decreases} from $T^*/T_F=0.7$ to  $T/T_F=0.2$   as density grows from 1  to $10\times 10^{11}$cm$^{-2}$. Therefore, the calculated $T^*(n)$ dependence does not agree with the experimental data shown in Fig.~3 of the main text, where  the inflection point {\it increases} with density from $T/T_F=0.09$ to $T/T_F=0.7$ in the same range of  densities.
In the latest version of the screening theory  \cite{das-hwang_PRB_2013}, all theoretical $\rho(T)$ curves have a positive curvature with no inflection point (ignoring the weak localization corrections which are beyond the scope of our paper and the above theories).

  We believe therefore, that currently no theory capture  the main feature of all experimental data for Si-MOSFETs, in the ``high temperature'' correlated regime, including the large $\rho(T)$ variations and
inflection in $\rho(T)$ whose temperature grows with density.
In the absence of a thorough microscopic theory, and having three clear prompts on the experimental side, we are eligible to suggest an empirical phenomenological form, that appears to fit the data with a minimal set of
parameters and links the transport features in a unified picture. The model attaches importance to the existence of a novel energy scale $T^*$, and presumes the existence of two parallel scattering channels.
Surely, the microscopic ground of the novel energy scale  requires separate study, that is beyond the framework of our paper.

 \subsection{Testing the empirical model}
Figure~4 of the main text shows comparison of the calculated $\rho(T)$ and $a_\sigma$ with experimental data.
The empirical dual resistivity model fits self-consistently both, the zero field resistivity $\rho(T)$  and the magnetoconductivity.
The  parameters of the fit are summarized in the Table below. The factor $\beta$ is an order of magnitude smaller than $\xi$, therefore, the second term in Eq.~(3) of the main text  becomes important only at high temperatures. The slope, $\alpha$, is almost constant, confirming the linear $\Delta(n)$ dependence (Eq.~(1) of the main text).

\begin{table}[h]
\caption{Summary of fitting parameters, corresponding to Fig.~4 and Eq.~(3) of the main text. $\rho_1$ and $\rho_D = \sigma_D^{-1}$ are in (k$\Omega/\Box $), density is in units of $10^{11}$\,cm$^{-2}$,  $n_c=0.88$, $\alpha$ is in K/$10^{11}$cm$^{-2}$.
}
\label{F0summarytable}
\begin{tabular}{|c|c|c|c|c|c|}
        \hline $n$ & $\rho_D$ & $\rho_1$ & $\alpha $ & $\beta$\,(K/T$^2$) & $\xi$\,(K$^2$/T$^2$)\\
		\hline 1.5 & 1268 & 14362 & 4.53 & -0.0160 & -0.08  \\
		\hline 1.996 & 901 & 9564 & 4.35 & -0.0080 & -0.09  \\
		\hline 2.5 & 662.2 & 6937 & 4.28 & -0.0043 & -0.11  \\
		\hline 3.25 & 501.5 & 5202 & 4.24 & -0.0019 & -0.15 \\
		\hline 5.252 & 336.14 & 3456.6 & 4.18 & -0.0005 & -0.19  \\
\hline
\end{tabular}
\end{table}

\subsection{On the role of phonons}
In 3D metals  any residual weak
temperature dependence in $\rho(T)$ originates from phonon
scattering which produces the  Bloch-Gruneisen
behavior, $\rho(T)=\rho_0 +\rho_1 T^5$, where the temperature-
independent contribution $\rho_0$ arises from short-range disorder
scattering and the  temperature dependence
(the second term) -- from  phonon scattering.  By contrast,
the temperature dependent transport in 2D metallic systems at low temperatures, besides  weak-localization effects,
 is dominated mostly by electron-impurity scattering dressed with electron-electron interaction effects (or on the complementary language -- by screened disorder scattering with temperature dependent screening).

%%%%%%%%%%%%%%%%%%%%%%%%%%%%%%%%%%%%%%%%%%%%%%
\begin{figure}[h]
\begin{center}
\includegraphics[width=400pt]{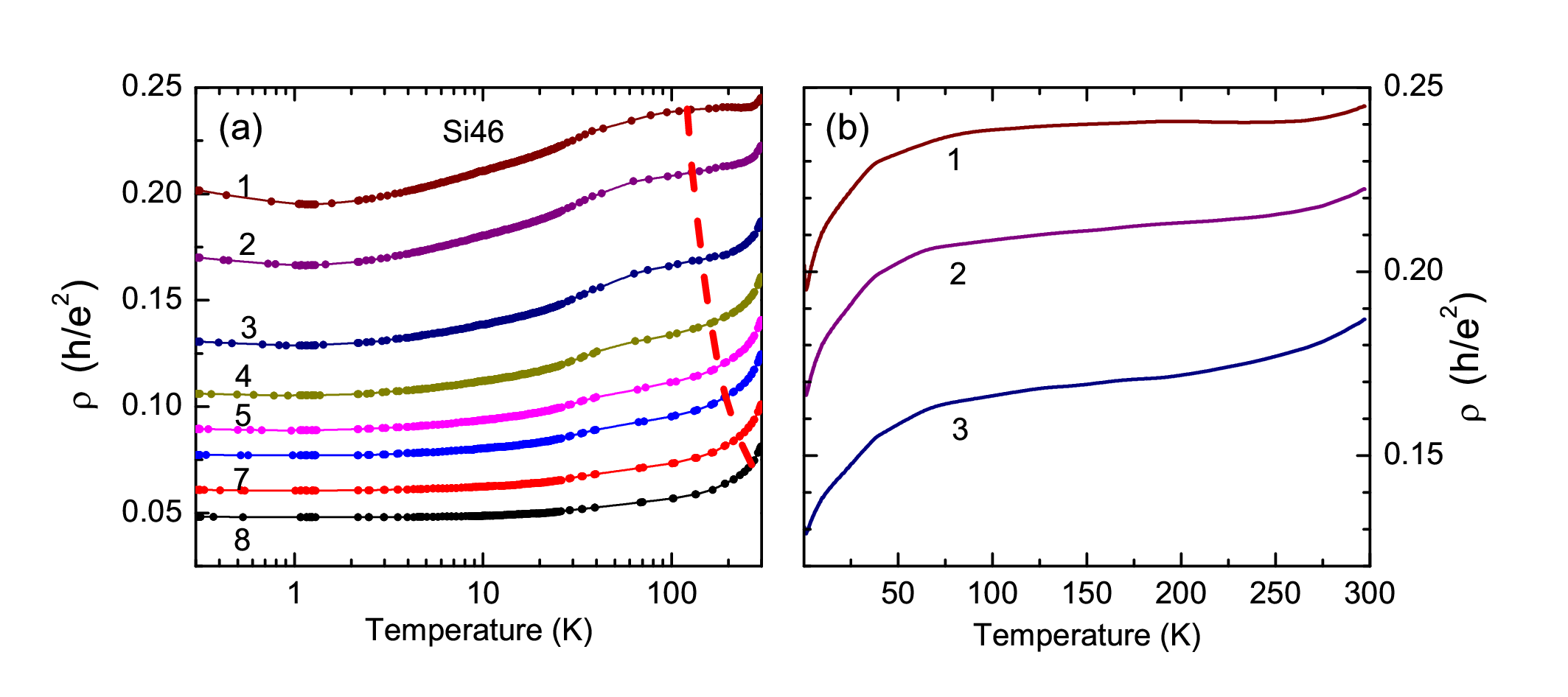}
\caption{(Color online)  Zero field temperature dependencies of  $\rho(T)$ for a low mobility sample in a wide range of temperatures. (a)  double-log, and (b) linear scale. Sample Si-46. Density: 1 -- 16.5, 2 -- 17.6, 3 -- 19.8, 4 -- 22, 5 -- 24.2, 6 -- 26.4, 7 -- 30.8, and 8 -- 36.3 $\times 10^{11}$cm$^{-2}$. Dashed line depicts $T_F$ for various densities.}
\label{fig:phonons}
\end{center}
\end{figure}
%%%%%%%%%%%%%%%%%%%%%%%%%%%%%%%%%%%%%%%%%%%%

 The interaction effects in transport  are proportional to $(T\tau)$ and in order to diminish them and to highlight the effect of phonons, we presented in Fig.~\ref{fig:phonons} the resistivity data for low mobility Si-MOS sample ($\tau$ is smaller by a factor of 10 than for the high mobility samples studied in the paper). One can see that
below about 2\,K, logarithmic quantum corrections  dominate (both, WL and interaction corrections) \cite{amp_pss_2000} (see Fig.~\ref{fig:phonons}a). For higher temperatures, up to the Fermi energy (dashed curve), ballistic interaction corrections (or temperature dependent screening) take over and cause $\rho(T)$ growing which flattens and then saturates as $T$ approaches $T_F$, due to nondegeneracy effects \cite{dassarma_2000}.  For temperatures higher than 100-- 200\,K, resistivity again starts growing, now due to electron-acoustic-phonon scattering. The monotonic $\rho \propto T$ dependence is a consequence of the amount of phonons excited at a given $T$.
In GaAs heterostructures, due to effective piezo-coupling the phonon scattering is rather strong \cite{dassarma_2000, zhou_2012}. For Si, the phonon scattering contribution to the overall scattering rate is much lower, because of the weaker electron-phonon coupling mechanism (that is the deformation potential for Si).

 To conclude, it is well-known that phonon scattering in Si-structures contributes essentially to the transport only in the vicinity of room temperature, and is irrelevant to the low-temperature transport. Both, the nondegeneracy and phonon scattering are  irrelevant to the inflection in $\rho(T)$ which happens at much lower temperatures than the onset of phonon scattering.
Nevertheless, to be on a safe side,  in the main text we analyze our data (kink in $\partial^2\sigma/\partial B^2$) and inflection (in $\rho(T)$) only in the temperature range (i) well below $E_F$  and below the $\rho(T)$ maximum in order to avoid the nondegeneracy effects, and (ii) always below 20K for the explored densities, where the  phonon contribution to the resistivity in Si-MOSFETs can be neglected with  1\% or better accuracy.